\documentclass[cameraready]{Interspeech}

\usepackage{hyperref}
\usepackage{multirow}
\title{DeSRPA: Decoupled Speech Role-Playing Agent via Inference-Time Intervention}

\author[affiliation={1}, orcid=0009-0008-6687-4460, correspondingauthor]{Wenqiu}{Tang}
\author[affiliation={2}, orcid=0009-0008-4550-4794]{Zhen}{Wan}
\author[affiliation={1}, orcid=0000-0002-3041-4330]{Takahiro}{Komamizu}
\author[affiliation={1}, orcid=0000-0003-3942-9296]{Ichiro}{Ide}
\address{
$^1$ Nagoya University, Nagoya, Aichi, Japan \
$^2$ National Institute of Informatics, Tokyo, Japan
}

\email{tang.wenqiu.t2@s.mail.nagoya-u.ac.jp, zhenwan@nii.ac.jp, taka-coma@acm.org, ide@i.nagoya-u.ac.jp}

\usepackage{comment}
\usepackage{enumitem}
\usepackage{makecell}

\begin{document}

\maketitle

\begin{abstract}
% While Large Language Models (LLMs) revolutionized text-based role-playing, they lack paralinguistic nuances like timbre and prosody essential for immersive interactions. Current Speech Role-Playing Agents (SRPAs) rely on End-to-End (E2E) fine-tuning. 

% However, this imposes a ``modality alignment tax'' degrading LLM reasoning, increasing costs, and reducing emotional interpretability. We propose a lightweight, decoupled SRPA framework for robust character adaptation via inference-time intervention instead of parameter updates. Our dual-level approach uses: 
% (1) \textbf{Internal Cognitive Steering}, injecting disentangled persona and style vectors into a frozen LLM, and (2) \textbf{External Expressive Rendering}, mapping dynamic emotion labels to style vectors. Experiments show our method outperforms both E2E baselines and cascaded pipelines in personality and emotion consistency, achieving competitive speech naturalness narrowing the gap with GPT-4o Audio.

While Large Language Models (LLMs) have revolutionized text-based role-playing, creating immersive Speech Role-Playing Agents (SRPAs) requires a seamless bridge between cognitive reasoning and paralinguistic nuances. Current SRPAs primarily rely on end-to-end (E2E) fine-tuning. However, this paradigm suffers from poor generalization to unseen characters due to its reliance on role-specific data, while imposing a ``modality alignment tax" that degrades intrinsic LLM reasoning capabilities.
We propose DeSRPA, an agentic framework for character role play via inference-time intervention on frozen backbones. DeSRPA employs a dual-level control vector mechanism 
\textbf{Internal Cognitive Steering} and
\textbf{External Expressive Rendering} to synchronize ``mind" and ``voice". Experiments on SpeechRole and OmniCharacter benchmarks demonstrate that DeSRPA significantly outperforms E2E baselines in personality and emotional consistency. It achieves high speech naturalness, narrowing the gap with proprietary models like GPT-4o Audio, while remaining a scalable and training-free paradigm. \footnote{Audio samples: \url{https://steeremo971-commits.github.io/emosteer-tts-demo/}}
\end{abstract}

\section{Introduction}
\label{sec:intro}

% --- 第一段：大背景与趋势 (Hook) ---
% 逻辑：RPAs 正在从纯文本走向语音多模态，强调情感表达的重要性。
Recent advancements in Large Language Models (LLMs) have revolutionized the development of Role-Playing Agents (RPAs), enabling them to simulate diverse personas with impressive linguistic fidelity~\cite{he-etal-2025-crab, wang-etal-2024-rolellm}. 
However, text-based interactions often lack the paralinguistic nuances, such as timbre, prosody, and emotion that are essential for truly immersive experiences~\cite{wang-etal-2025-benchmarking-contextual}. 
To bridge this gap, Speech Role-Playing Agents (SRPAs) have emerged as a promising paradigm, aiming to unify cognitive reasoning with acoustic expression to create realistic voice-based characters.

% --- 第二段：现有方法的痛点 (Gap) ---
% 逻辑：指出主流 E2E 方法（Omni 系列）依赖大量数据 SFT，导致“数据昂贵”和“控制解耦难”的问题。

Despite this progress, state-of-the-art SRPAs ~\cite{jiang2026speechrolelargescaledatasetbenchmark, zhang-etal-2025-omnicharacter} predominantly rely on End-to-End (E2E) supervised fine-tuning. This paradigm faces a critical ``generalization trap": the heavy reliance on role-specific datasets limits the model's scalability to unseen characters and incurs prohibitive adaptation costs. Furthermore, joint audio-text modeling often imposes a ``modality alignment tax", where the pursuit of acoustic alignment inadvertently degrades the LLM’s intrinsic reasoning and persona-consistency capabilities. While traditional cascaded pipelines preserve LLM's capability, they suffer from semantic-acoustic misalignment ; the TTS module acts merely as a detached acoustic renderer, struggling to express dynamic emotional contexts and character-aware vocal nuances.

% --- 第三段：你的核心方法 (Solution) ---
% 逻辑：提出“解耦”和“控制向量”方案。LLM 控制人格与情绪标签，TTS 负责执行，实现轻量化。
To address these challenges, we propose the Decoupled Speech Role-Playing Agent (DeSRPA). Building on the agentic pipeline to preserve LLM reasoning, DeSRPA introduces a novel \textit{inference-time intervention} mechanism for lightweight character adaptation without parameter updates. It steers agent behavior via two synchronized levels: 
(1) \textbf{Internal Cognitive Steering:} This stage injects disentangled control vectors that represent persona, context, and linguistic style directly into the LLM’s residual stream to shape its internal reasoning. 
(2) \textbf{External Expressive Rendering:} Guided by stage 1, the LLM then generates responses corresponding explicit emotion tags that naturally reflect the character's unique personality and stance. These tags are then mapped to acoustic emotional control vectors to shape precise emotions in a TTS module. 

% --- 第四段：贡献总结 (Contributions) ---
% 逻辑：总结三大贡献：框架创新、双重一致性、实验SOTA。
The contributions of this work are summarized as follows:
\begin{itemize}
    \item We propose a training-free framework DeSRPA that leverages inference-time intervention for character adaptation. This approach achieves robust role-playing capabilities without the computational cost of E2E fine-tuning or the risk of reasoning degradation.
    \item We introduce a synchronization strategy that aligns internal cognitive steering with external expressive rendering. By mapping persona-driven intent to acoustic control vectors, the system ensures high consistency between character personality and paralinguistic expression.
    \item DeSRPA outperforms state-of-the-art E2E baselines in personality and emotional consistency. Furthermore, it demonstrates superior scalability in open-domain scenarios while narrowing the naturalness gap with proprietary models.
\end{itemize}
\begin{figure*}[t]
  \centering
  \includegraphics[width=0.9\textwidth]{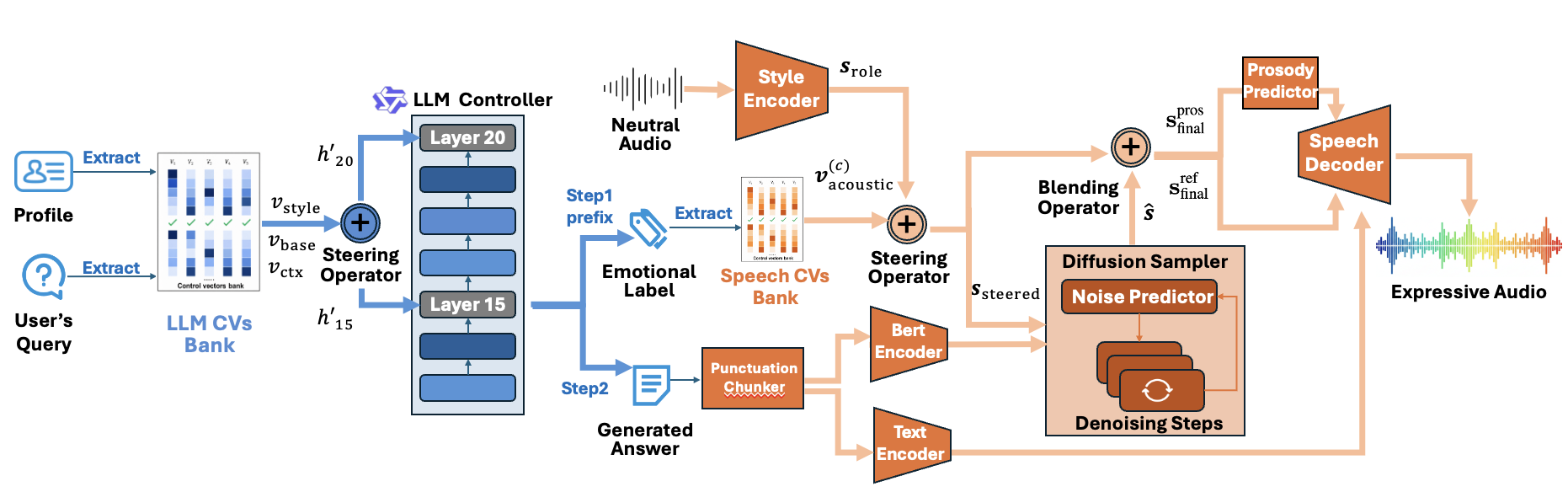}
  \caption{Proposed Decoupled Speech Role-Playing Agent (DeSRPA) framework. A \textbf{Frozen LLM Controller} (left) steers a \textbf{Frozen StyleTTS~2~\cite{NEURIPS2023_3eaad2a0}} (right) via \textit{Inference-Time Control Vectors}, injecting personality and acoustic styles directly without parameter updates.}
  \label{fig:framework}
\end{figure*}

\section{Related Work}
\label{sec:relwork}

While LLM-based Role-Playing Agents like RoleLLM~\cite{wang-etal-2024-rolellm} and CharacterGLM~\cite{zhou-etal-2024-characterglm} achieve high linguistic fidelity, cascaded LLM-TTS pipelines introduce \textit{semantic-acoustic misalignment}~\cite{zhang2023speechgptempoweringlargelanguage}. The intermediate textual bottleneck drops essential emotional reasoning, preventing TTS from rendering paralinguistic nuances. Furthermore, existing systems relying on discrete style labels fail to provide fine-grained control over emotional intensity.

To mitigate these cascading errors, End-to-End (E2E) models
% like GPT-4o~\cite{openai2024gpt4o}, SpeechGPT~\cite{zhang2023speechgptempoweringlargelanguage}, and LLaMA-Omni~\cite{fang-etal-2025-llama} 
integrate speech modalities directly. However, adapting them for role-playing requires resource-intensive Supervised Fine-Tuning (SFT) on massive datasets like SpeechRole~\cite{jiang2026speechrolelargescaledatasetbenchmark} and VoxRole~\cite{wu2025voxrolecomprehensivebenchmarkevaluating}, which limits open-domain scalability.

Alternatively, inference-time intervention steers outputs without extensive retraining. For instance, Representation Engineering (RepE)~\cite{zou2025representationengineeringtopdownapproach} successfully manipulates LLM behavior via internal hidden states, while speech models like EmoSphere-TTS~\cite{Cho_2025} and EmoSteer-TTS~\cite{xie2025emosteerttsfinegrainedtrainingfreeemotioncontrollable} enable training-free emotion control. Nevertheless, these methods often rely on complex geometric mappings or heuristic searches. To address the aforementioned limitations across both cascaded and E2E systems, our decoupled framework introduces a dual-level vector injection mechanism. By leveraging the explicitly disentangled style space of StyleTTS~2~\cite{NEURIPS2023_3eaad2a0}, we achieve a direct, training-free inference-time intervention. This approach bypasses costly audio SFT and synchronizes LLM text generation with TTS acoustic expression, ensuring fine-grained, holistic character consistency.

\section{Methodology}
\label{sec:method}

As illustrated in Fig~\ref{fig:framework}, the proposed framework bridges the gap between cognitive reasoning and acoustic expression via a dual-level control mechanism that operates on two frozen backbones:

\textbf{Internal Cognitive Steering:} DeSRPA utilizes a frozen LLM, Qwen3-4B \cite{qwen3}, as the cognitive brain. Instead of updating weights, it injects disentangled \textit{Personality Vectors} and \textit{Language Feature Vectors} into the inference stream. This steers the model to generate persona-aligned text with personality-based emotional labels.

\textbf{External Expressive Rendering:} For vocal synthesis, DeSRPA employs a frozen StyleTTS~2 model~\cite{NEURIPS2023_3eaad2a0}. LLM-predicted emotional labels map to acoustic control vectors within a \textit{Vector Bank}. This vector is applied via a \textit{Steering Operation} to the base style embedding. The modulated style representation then conditions the Diffusion Sampler, guiding the generation of expressive, character-consistent audio.

\subsection{Internal Cognitive Steering (LLM Controller)}
To adapt the agent's behavior without parameter updates, DeSRPA intervenes in the LLM's residual stream using layer-specific control vectors.

\subsubsection{Cognitive Vectors via Sparse Autoencoders}
DeSRPA controls the LLM's personality via Sparse AutoEncoders (SAEs) \cite{he-etal-2025-sae}. This method optimizes a sparse vector $\mathbf{v}$ to shift latent representations $\mathbf{z}$ (where $\mathbf{z}^{\prime} = \mathbf{z} + \mathbf{v}$) toward a target centroid $\boldsymbol{\mu}^{+}$ and away from an opposing $\boldsymbol{\mu}^{-}$ by minimizing:
\begin{equation}
    \mathcal{L}_{\mathbf{steer}} = \Vert \mathbf{z}^{\prime} - \boldsymbol{\mu}^{+} \Vert_{2}^{2} - \Vert \mathbf{z}^{\prime} - \boldsymbol{\mu}^{-} \Vert_{2}^{2} + \mathcal{L}_{\mathbf{LM}} + \lambda \Vert \mathbf{v} \Vert_{1},
\end{equation}
where $\mathcal{L}_{\mathbf{LM}}$ preserves the generation quality and $\lambda \Vert \mathbf{v}_{} \Vert_{1}$ enforces the sparsity. 
Since this is a layerwise process, in our preliminary work and guided by findings \cite{zhong2024englishcentricllmslanguagemultilingual}, we found that mid-layers process core semantics while deeper layers govern surface styles, we train three functional types of vectors at specific depths:
\begin{itemize}
    \item \textbf{Personality Base ($\mathbf{v}_{\text{base}}$) \& Contextual Activation ($\mathbf{v}_{\text{ctx}}$)}: These are optimized at Layer 15 to modulate core identity and situational reasoning.
    \item \textbf{Linguistic Style ($\mathbf{v}_{\text{style}}$)}: This is optimized at Layer 20 to capture character-specific phrasing and idiolect.
\end{itemize}
Extending beyond the standard Big Five dimensions~\cite{sun-etal-2025-personality}, we train vectors across 30 fine-grained facets using a 15k dataset~\cite{tang2026facetlevelpersonacontroltraitactivated} to formulate three types of control vectors: $\mathbf{v}_{\mathbf{base}}$, $\mathbf{v}_{\mathbf{style}}$, and $\mathbf{v}_{\mathbf{ctx}}$.

\subsubsection{Dynamic Inference-Time Intervention}
During inference, DeSRPA functions as an active agent by dynamically scaling these vectors according to the character's profile and current query. The modified hidden states are computed as:
\begin{equation}
    \mathbf{h}_{15}' = \mathbf{h}_{15} + w_{b} \mathbf{v}_{\text{base}} + w_{c} \mathbf{v}_{\text{ctx}}, \quad \mathbf{h}_{20}' = \mathbf{h}_{20} + w_{s} \mathbf{v}_{\text{style}},
\end{equation}
where $\mathbf{h}$ represents the original residual stream and $w_{b}, w_{c}, w_{s}$ are scaling coefficients grounded in objective personality metrics from PDB. These coefficients are refined via Human-LLM collaborative annotation, achieving strong inter-rater agreement (Pearson's $r = 0.82$). Based on Trait Activation Theory~\cite{tett2003personality}, $\mathbf{v}_{\text{ctx}}$ is dynamically activated by the LLM to ensure context-appropriate persona expression.

\subsection{External Expressive Rendering (TTS Module)}
To bridge the gap between cognitive intent and vocal execution without training, DeSRPA utilizes a frozen StyleTTS 2 backbone. This module renders expressive speech by modulating its latent style space using emotion-aware control vectors.

\subsubsection{Acoustic Vectors via Style Subtraction}
To bridge the gap between textual persona and acoustic expression, we construct control vectors that define the ``outer'' voice. We leverage two  multi-speaker, parallel emotional speech datasets: the Emotional Speech Database (ESD)~\cite{zhou2022emotionalvoiceconversiontheory} and Crowd-sourced Emotional Multimodal Actors Dataset (CREMA-D)~\cite{6849440}. Together, they provide diverse speakers and a wide range of emotions for robust CV' training.

\noindent \textbf{Data Filtering:} To ensure the purity of the representations, we apply a rigorous filtering pipeline. For each target emotion category $c \in \{\text{neutral}, \text{angry}, \text{happy}, \text{sad}, \text{surprise}, \text{disgust}, \text{fear}\}$, specifically, we evaluate the generated samples using \textit{Emo2Vec}~\cite{xu-etal-2018-emo2vec} scores ($> 0.90$) to ensure high emotional intensity, and \textit{Silence Rate} ($< 20\%$) to guarantee audio stability. We then select the highest quality $N (=300)$ samples per emotion based on these criteria.

\noindent \textbf{Vector Extraction:} We leverage the pre-trained Style Encoder and Predictor Encoder of StyleTTS~2~\cite{NEURIPS2023_3eaad2a0} to extract a unified style representation $S(x) = [\mathbf{r}_s; \mathbf{r}_p] \in \mathbb{R}^{256}$ for input audio $x$.

\noindent \textbf{Difference Vector Computation:} Directly injecting emotional audio embeddings often entangles the target emotion with the reference speaker's identity~\cite{10.1109/TASLP.2022.3164181}. To mitigate this, we propose a \textit{style subtraction} method. We compute the difference between the mean embedding of the target emotion and the mean embedding of the neutral state from the same parallel corpus:
\begin{equation}
    \mathbf{v}_{\text{acoustic}}^{(c)} = \frac{1}{N} \sum_{i=1}^{N} S(x_{i}^{(c)}) - \frac{1}{N} \sum_{i=1}^{N} S(x_{i}^{(n)})
\end{equation}
where $x^{(c)}$ and $x^{(n)}$ denote the filtered samples for emotion $c$ and the neutral category, respectively. The resulting vector $\mathbf{v}_{\text{acoustic}}^{(c)}$ captures the \textit{direction} of the emotional shift in the latent space, independent of speaker identity (see SIM in Table~\ref{table1}, where all values $> 0.85$).

\subsubsection{Acoustic Injection}
For speech synthesis, we propose a \textit{Dual-Path Fusion Strategy} to balance speaker identity retention with emotional expressiveness, consisting of four steps:
\begin{enumerate}
    \item \textbf{Style Extraction:} Following StyleTTS~2~\cite{NEURIPS2023_3eaad2a0}, speaking style is encoded as a 256-dimensional vector $\mathbf{s} = [\mathbf{s}^{\text{ref}}; \mathbf{s}^{\text{pros}}] \in \mathbb{R}^{256}$, where $\mathbf{s}^{\text{ref}} \in \mathbb{R}^{128}$ is extracted by the Style Encoder to capture speaker timbre, and $\mathbf{s}^{\text{pros}} \in \mathbb{R}^{128}$ is extracted by the Style Predictor's encoder to capture prosodic characteristics. At inference time, we establish our baseline by extracting a role-specific reference style $\mathbf{s}_{\text{role}}$ from a target neutral utterance.

    \item \textbf{Latent Steering:} We inject emotional expressiveness by steering the latent representation via vector arithmetic as:
    \begin{equation}
      \mathbf{s}_{\text{steered}} = \mathbf{s}_{\text{role}} + \tau \mathbf{v}_{\text{acoustic}}^{(c)},
      \label{eq:steering}
    \end{equation}
    where the emotion intensity scalar $\tau \in [0.5, 2.5]$ is determined by the product of the LLM-inferred emotion label weight and the annotated intensity score. The range boundaries correspond to subtle emotional onset ($\tau=0.5$) and peak expressiveness ($\tau=2.5$), as identified through preliminary coefficient sweeping.

    \item \textbf{Diffusion-based Refinement:} To incorporate text-aligned prosodic constraints, $\mathbf{s}_{\text{steered}}$ is passed as the reference conditioning signal into the diffusion-based Style Predictor, yielding a text-coherent predicted style $\hat{\mathbf{s}}$.

    \item \textbf{Dual-Path Interpolation:} To balance emotional intensity with natural prosody, we perform a final interpolation between the predicted ($\hat{\mathbf{s}}$) and steered ($\mathbf{s}_{\text{steered}}$) styles:
    \begin{align}
      \mathbf{s}^{\text{ref}}_{\text{final}} &= (1 - \rho) \hat{\mathbf{s}}^{\text{ref}} + \rho \mathbf{s}^{\text{ref}}_{\text{steered}}, \label{eq:blend_ref} \\
      \mathbf{s}^{\text{pros}}_{\text{final}} &= (1 - \eta)\hat{\mathbf{s}}^{\text{pros}} + \eta \mathbf{s}^{\text{pros}}_{\text{steered}}. \label{eq:blend_pros}
    \end{align}
\end{enumerate}
We selected $\rho = 0.8$ as it consistently maintained Speaker Similarity (SIM) above the speaker verification threshold. For $\eta$, we observed that values below $0.5$ yielded insufficient emotional expressiveness in Emotion Execution Accuracy (EEA), while values above $0.5$ produced diminishing returns in EEA.
\section{Experiment}

\begin{table*}[t]
\centering
\small % 缩小整体字体以节省空间
\setlength{\tabcolsep}{4.5pt} % 缩小列间距，使表格更加紧凑
\caption{Comprehensive evaluation on the SpeechRole \cite{jiang2026speechrolelargescaledatasetbenchmark} dataset, among open-source models, the best results are highlighted in \textbf{bold}, and the second-best are \underline{underlined}.}
\label{table1}
\begin{tabular}{lccccccccc}
\toprule
\multirow{2}{*}{\textbf{Metric}}
    & \multicolumn{4}{c}{\textbf{Open}} & \multicolumn{3}{c}{\textbf{Ablation (DeSRPA $-$ CVs)}}  & \multicolumn{2}{c}{\textbf{Proprietary}} \\
\cmidrule(lr){2-5} \cmidrule(lr){6-8} \cmidrule(lr){9-10}
    & \makecell{Qwen2.5 \\ \cite{xu2025qwen25omnitechnicalreport}}
    & \makecell{LLaMA \\ \cite{fang-etal-2025-llama}}
    & \makecell{SpeechRole\\ \cite{jiang2026speechrolelargescaledatasetbenchmark}}
    & \makecell{DeSRPA \\ (Ours)}
    & \makecell{w/o \\ Both}
    & \makecell{w/o \\ LLM}
    & \makecell{w/o \\ Speech}
    & \makecell{GPT-4o \\ \cite{openai2024gpt4o}}
    & \makecell{AliCloud \\ \cite{alibabacloud2026}}
    \\ \midrule

\multicolumn{10}{c}{\textbf{Objective Metrics}} \\
TTFA (ms) ($\downarrow$) & \underline{274}      & \textbf{226} & 389    & 577    & 561    & 573      & 569     & 320    & 872    \\
SIM ($\uparrow$) & $<0.80$ & $<0.80$ & $<0.80$ & \textbf{0.886} & 0.919 & 0.905 & 0.892 & $<0.80$ & 0.859 \\
EEA ($\uparrow$)         &  \underline{0.453}      & 0.397      & 0.433     & \textbf{0.701} & 0.537  & 0.677  & 0.549  & 0.501      & 0.694  \\
WER (\%) ($\downarrow$)  & \textbf{0.98} & \underline{2.21}   & 5.31   & 2.63   & 2.49   & 2.64   & 2.52   & 2.03   & 1.74   \\
\addlinespace % 仅用一点微小的留白代替粗重的 \midrule 来分隔
\multicolumn{10}{c}{\textbf{Multimodal Judge Evaluation}} \\
Inst. Adherence       & 0.5127 & 0.7808 & \underline{0.8203} & \textbf{0.8790} & 0.8745 & 0.8810 & 0.8799 & 0.9137 & 0.8986 \\
Speech Fluency        & 0.6714 & \textbf{0.8795} & \underline{0.8745} & 0.8741 & 0.8922 & 0.8891 & 0.8846 & 0.9329 & 0.8706 \\
Conv. Coherence       & 0.6326 & 0.8607 & \underline{0.9316} & \textbf{0.9506} & 0.9548 & 0.9245 & 0.9351 & 0.9983 & 0.9563 \\
Speech Naturalness    & 0.6474 & \underline{0.7864} & 0.7838 & \textbf{0.8147} & 0.8155 & 0.8152 & 0.8149 & 0.9079 & 0.8177 \\
Prosodic Consist.     & 0.4763 & \underline{0.6620} & 0.6505 & \textbf{0.7958} & 0.6936 & 0.7412 & 0.7186 & 0.8046 & 0.7743 \\
Emotion Approp.       & 0.4793 & 0.6427 & \underline{0.6913} & \textbf{0.8160} & 0.6874 & 0.7568 & 0.7245 & 0.8341 & 0.7872 \\
Personality Const.    & 0.3931 & \underline{0.6415} & 0.6122 & \textbf{0.7615} & 0.7167 & 0.7235 & 0.7592 & 0.8018 & 0.7402 \\
Knowledge Const.      & 0.5907 & 0.7077 & \textbf{0.8334}    & \underline{0.8116} & 0.7828 & 0.7743 & 0.8074 & 0.8910 & 0.8405 \\ \midrule
Mean (Automated Eval)     & 0.5504 & 0.7452 & \underline{0.7747} & \textbf{0.8379} & 0.8022 & 0.8120 & 0.8168 & 0.8862 & 0.8356 \\
\bottomrule
\end{tabular}
\vspace{-10pt} % 如果页面底部还是有些挤，可以保留这个命令微调
\end{table*}

\subsection{Experimental Settings}

To evaluate character fidelity under both plot-based and open-domain interaction scenarios, we utilize two distinct datasets.
SpeechRole-Data test split~\cite{jiang2026speechrolelargescaledatasetbenchmark} serves as our primary bench mark for plot-based evaluation, where queries have verifiable answers anchored to established source material. We curate a subset of 72 English characters from movies and TV series, each engaging in an average of 10 dialogue turns, resulting in 372 evaluated responses ($N=372$).
Complementing this, OmniCharacter-10K test split~\cite{zhang-etal-2025-omnicharacter} covers open-domain scenarios using 10 characters from the RPG \textit{Genshin Impact}, where queries extend beyond the source narrative and require the model to extrapolate character responses to novel topics.

%\subsubsection{Evaluation Metrics.}
We adopt a hybrid evaluation protocol:

\noindent\textbf{Multimodal Judge Evaluation:} Following~\cite{jiang2026speechrolelargescaledatasetbenchmark}, we employ the multimodal model Gemini 2.5 Pro~\cite{comanici2025gemini25pushingfrontier} to assess conversational capabilities on SpeechRole. Conditioned on identical profiles, references, and prompts, the judge performs pairwise relative scoring against ground truths across eight dimensions: \textit{instruction adherence}, \textit{fluency}, \textit{coherence}, \textit{naturalness}, \textit{prosody}, \textit{emotion}, \textit{personality}, and \textit{knowledge}.
    
\noindent\textbf{Objective Metrics:} To rigorously quantify acoustic fidelity and responsiveness, we evaluate models on the SpeechRole dataset using four metrics: \textit{Time-to-First-Audio (TTFA)} measures streaming latency from user input to the first audio chunk~\cite{zhang-etal-2025-omnicharacter}; \textit{Speaker Similarity (SIM)} evaluates timbre consistency via cosine similarity of WaveLM~\cite{Chen_2022} speaker embeddings against references; \textit{Emotion Execution Accuracy (EEA)} utilizes emotion2vec~\cite{ma-etal-2024-emotion2vec} to assess consistency between the predicted emotion labels and emotions rendered in the generated speech; and \textit{Word Error Rate (WER)} assesses speech intelligibility via ASR transcriptions~\cite{zhang-etal-2025-omnicharacter}. 

\noindent\textbf{Human Evaluation:} To capture paralinguistic nuances and interactive quality in open-domain scenarios, we conduct human evaluations on the OmniCharacter dataset. Expert evaluations assess six dimensions across three core aspects: \textit{speech quality} (fluency, clarity), \textit{persona expressiveness} (emotion, consistency), and \textit{interactive experience} (appropriateness, immersion)~\cite{zhang-etal-2025-omnicharacter}.

%\subsubsection{Baselines.}

We evaluate our method against end-to-end (E2E) models and cascaded pipelines. E2E baselines include LLaMA-Omni~\cite{fang-etal-2025-llama}, Qwen2.5-Omni~\cite{xu2025qwen25omnitechnicalreport}) and fine-tuned role-playing specialists (SpeechRole~\cite{jiang2026speechrolelargescaledatasetbenchmark}, Omni Character~\cite{zhang-etal-2025-omnicharacter}). For cascaded baselines,  we include baselines that are ablations without control vectors based on our proposal. We also compared with proprietary API-based settings, such E2E models (GPT-4o Audio~\cite{openai2024gpt4o}) and cascaded AliCloud pipeline (\texttt{qwen-plus-character} with CosyVoice3~\cite{du2025cosyvoice3inthewildspeech}).

\subsection{Experiment Result}

\textbf{Multimodal Judge Evaluation}:
Table~\ref{table1} evaluates 72 roles. Among open-source models, our proposed method achieved the highest mean score of \textbf{0.8379}, substantially outperforming baselines like SpeechRole (0.7747) and LLaMA (0.7452). Furthermore, it performed comparably to proprietary systems, ranking second overall behind GPT-4o~\cite{openai2024gpt4o} (\textbf{0.8862}). Compared to the AliCloud baseline~\cite{alibabacloud2026}, which excels in content generation, the proposed method showed superior paralinguistic control. Specifically, it achieved  higher scores in \textit{Prosodic Consistency} (0.7958 vs. 0.7743) and \textit{Emotion Appropriateness} (0.8160 vs. 0.7872), highlighting its effectiveness in preserving acoustic and emotional fidelity.

\noindent\textbf{Objective Metrics Analysis.} Most importantly, DeSRPA ranked first in EEA with a score of 0.701. This confirmed that our method successfully aligned the LLM's predicted semantic emotion labels with the final TTS acoustic execution, providing expressive and controllable speech generation. In terms of SIM, because most end-to-end (E2E) models lack the capability to imitate specific voice timbres, their scores fell below 0.80. In contrast, our method achieved a strong SIM of 0.886. Integrating control vectors introduced an acceptable trade-off: compared to the ablation baseline, it only marginally increased WER to 2.63\% and slightly lowered SIM to this 0.886 level. Regarding responsiveness, DeSRPA achieved a TTFA of 577 ms. While the architectural cost naturally placed its latency higher than fully E2E models like LLaMA-Omni (226 ms)~\cite{fang-etal-2025-llama}, its lightweight design significantly outperformed heavier pipelines like the AliCloud API~\cite{alibabacloud2026}.

\noindent\textbf{Ablation study}:
To evaluate the contributions of the proposed modules, we conducted an ablation study (Table~\ref{table1}). While the pure baseline (w/o Both CVs) maintained strong general capabilities, it lacked character alignment. Specifically, LLM CVs govern textual identity; removing them degraded \textit{Personality Const.} (0.7615 $\rightarrow$ 0.7235) and \textit{Knowledge Const.} (0.8116 $\rightarrow$ 0.7743). Conversely, since Speech CVs control acoustic expressiveness, ablating them not only impaired subjective \textit{Prosodic Consistency} (0.7958 $\rightarrow$ 0.7186) and \textit{Emotion Approp.} (0.8160 $\rightarrow$ 0.7245), but also caused a drastic drop in objective \textit{Emotion Execution Accuracy} (EEA) from \textbf{0.701} to 0.549. As noted earlier, we observed an inherent trade-off: injecting CVs slightly perturbs the unguided baseline's hidden states, causing marginal drops in \textit{Fluency}, \textit{Naturalness}, \textit{Adherence}, and objective SIM/WER. Nevertheless, integrating both CVs is essential for bridging text--audio alignment, achieving the highest overall role-playing fidelity (Mean: \textbf{0.8379}).

\begin{table}[t]
  \caption{Performance comparison with the state-of-the-art methods on 10 English characters from the OmniCharacter-10K test split~\cite{zhang-etal-2025-omnicharacter}. Human evaluation scores are assessed on a 10-point Likert scale by 6 experts across six dimensions: fluency (Flu.), consistency (Cons.), emotional expression (Emo.), clarity (Cla.), appropriateness (App.), and immersion (Imm.).}
  \label{tab:performance_comparison}
  \centering
  % 适当缩小列间距，确保表格完美塞进单栏
  \setlength{\tabcolsep}{3.5pt} 
  % @{} 用于强制去除表格最左和最右的隐形边距
  \begin{tabular}{@{} l c c c c c c @{}}
    \toprule
    \textbf{Models} & \textbf{Flu.} & \textbf{Cons.} & \textbf{Emo.} & \textbf{Cla.} & \textbf{App.} & \textbf{Imm.} \\
    \midrule
    LLaMA-Omni~\cite{fang-etal-2025-llama}    & 6.88 & 4.27 & 3.44 & 6.69 & 4.78 & 4.68 \\
    OmniCharacter~\cite{zhang-etal-2025-omnicharacter} & \underline{7.97} & \textbf{6.84} &\underline{6.23} & \underline{7.88} & \textbf{5.63} & \textbf{8.52} \\
    \midrule
    Proposed          & \textbf{8.70} &\underline{ 6.07} & \textbf{7.41} & \textbf{9.11} & \underline{5.54} & \underline{7.44} \\
    \bottomrule
  \end{tabular}
\end{table}

\noindent\textbf{Human Evaluation}:
As shown in Table~\ref{tab:performance_comparison}, the proposed method achieved the highest scores in \textit{Fluency} (8.70), \textit{Clarity} (9.11), and \textit{Emotional Expression} (7.41), demonstrating superior acoustic quality and emotional resonance. The fixed-voice model LLaMA-Omni~\cite{fang-etal-2025-llama} naturally scores poorly in \textit{Consistency} and \textit{Immersion}. While our method supports dynamic voice adaptation like OmniCharacter~\cite{zhang-etal-2025-omnicharacter}, our \textit{Consistency} (6.07) and \textit{Immersion} (7.44) were comparatively lower. This discrepancy arises because OmniCharacter-10K~\cite{zhang-etal-2025-omnicharacter} features highly stylized anime personae with exaggerated prosody. These Out-of-Distribution traits challenge the TTS module, which is optimized for natural human speech.

\section{Conclusion}
We propose DeSRPA, a method for high-fidelity character adaptation via lightweight inference-time intervention that bypasses resource-intensive E2E fine-tuning. By injecting dual-level control vectors into frozen LLM and StyleTTS~2~\cite{NEURIPS2023_3eaad2a0} backbones, DeSRPA tightly aligns cognitive personae with acoustic expression. Experiments demonstrate that DeSRPA significantly outperforms open-source E2E baselines in personality consistency and emotion appropriateness while achieving competitive speech naturalness, narrowing the gap with GPT-4o Audio~\cite{openai2024gpt4o}. This validates tuning-free deep style control, providing a scalable paradigm for open-domain speech systems.

\newpage
\section{Acknowledgments}
This work was supported by JSPS KAKENHI JP25K00161

\section{Generative AI Use Disclosure}

Generative AI tools were used only for language polishing, grammar correction, and improving the clarity of this work. All arguments, analysis, and conclusions are my own. AI-generated suggestions were reviewed and edited before inclusion.

\bibliographystyle{IEEEtran}
\bibliography{mybib}

\begin{thebibliography}{10}
\providecommand{\url}[1]{#1}
\csname url@samestyle\endcsname
\providecommand{\newblock}{\relax}
\providecommand{\bibinfo}[2]{#2}
\providecommand{\BIBentrySTDinterwordspacing}{\spaceskip=0pt\relax}
\providecommand{\BIBentryALTinterwordstretchfactor}{4}
\providecommand{\BIBentryALTinterwordspacing}{\spaceskip=\fontdimen2\font plus
\BIBentryALTinterwordstretchfactor\fontdimen3\font minus \fontdimen4\font\relax}
\providecommand{\BIBforeignlanguage}[2]{{%
\expandafter\ifx\csname l@#1\endcsname\relax
\typeout{** WARNING: IEEEtran.bst: No hyphenation pattern has been}%
\typeout{** loaded for the language `#1'. Using the pattern for}%
\typeout{** the default language instead.}%
\else
\language=\csname l@#1\endcsname
\fi
#2}}
\providecommand{\BIBdecl}{\relax}
\BIBdecl

\bibitem{he-etal-2025-crab}
K.~He, Y.~Huang, W.~Wang, D.~Ran, D.~Sheng \emph{et~al.}, ``{Crab}: A novel configurable role-playing {LLM} with assessing benchmark,'' in \emph{Proceedings of the 63rd Annual Meeting of the Association for Computational Linguistics}, vol.~1, Vienna, Austria, Jul. 2025, pp. 15\,030--15\,052.

\bibitem{wang-etal-2024-rolellm}
N.~Wang, Z.~Peng, H.~Que, J.~Liu, W.~Zhou \emph{et~al.}, ``{R}ole{LLM}: Benchmarking, eliciting, and enhancing role-playing abilities of large language models,'' in \emph{Findings of the Association for Computational Linguistics: ACL 2024}, Bangkok, Thailand, Aug. 2024, pp. 14\,743--14\,777.

\bibitem{wang-etal-2025-benchmarking-contextual}
Q.~Wang, H.~B. Sailor, T.~Liu, W.~Zhang, M.~Huzaifah \emph{et~al.}, ``Benchmarking contextual and paralinguistic reasoning in speech-{LLM}s: A case study with in-the-wild data,'' in \emph{Findings of the Association for Computational Linguistics: EMNLP 2025}, Suzhou, China, Nov. 2025, pp. 14\,133--14\,148.

\bibitem{jiang2026speechrolelargescaledatasetbenchmark}
C.~Jiang, J.~Sun, Y.~Cao, J.~Zhuang, H.~Li \emph{et~al.}, ``{SpeechRole}: A large-scale dataset and benchmark for evaluating speech role-playing agents,'' \emph{Computing Research Repository, arXiv Preprint, \textup{arXiv:2508.02013}}, Aug 2025.

\bibitem{zhang-etal-2025-omnicharacter}
H.~Zhang, R.~Luo, X.~Liu, Y.~Wu, T.-E. Lin, and P.~Zeng, ``{O}mni{C}haracter: Towards immersive role-playing agents with seamless speech--language personality interaction,'' in \emph{Proceedings of the 63rd Annual Meeting of the Association for Computational Linguistics}, Vienna, Austria, Jul. 2025, pp. 26\,318--26\,331.

\bibitem{NEURIPS2023_3eaad2a0}
Y.~A. Li, C.~Han, V.~Raghavan, G.~Mischler, and N.~Mesgarani, ``Styletts 2: Towards human-level text-to-speech through style diffusion and adversarial training with large speech language models,'' in \emph{Advances in Neural Information Processing Systems}, A.~Oh, T.~Naumann, A.~Globerson, K.~Saenko, M.~Hardt, and S.~Levine, Eds., vol.~36, 2023, pp. 19\,594--19\,621.

\bibitem{zhou-etal-2024-characterglm}
J.~Zhou, Z.~Chen, D.~Wan, B.~Wen, Y.~Song \emph{et~al.}, ``{C}haracter{GLM}: Customizing social characters with large language models,'' in \emph{Proceedings of the 2024 Conference on Empirical Methods in Natural Language Processing: Industry Track}, Miami, FL, USA, Nov. 2024, pp. 1457--1476.

\bibitem{zhang2023speechgptempoweringlargelanguage}
D.~Zhang, S.~Li, X.~Zhang, J.~Zhan, P.~Wang, Y.~Zhou, and X.~Qiu, ``{S}peech{GPT}: Empowering large language models with intrinsic cross-modal conversational abilities,'' \emph{Findings of the Association for Computational Linguistics: EMNLP 2023}, pp. 15\,757--15\,773, Dec. 2023.

\bibitem{wu2025voxrolecomprehensivebenchmarkevaluating}
W.~Wu, L.~Cao, X.~Wu, Z.~Lin, R.~Niu \emph{et~al.}, ``{VoxRole}: A comprehensive benchmark for evaluating speech-based role-playing agents,'' \emph{Computing Research Repository, arXiv Preprint, \textup{arXiv:2509.03940}}, Sep. 2025.

\bibitem{zou2025representationengineeringtopdownapproach}
A.~Zou, L.~Phan, S.~Chen, J.~Campbell, P.~Guo \emph{et~al.}, ``Representation engineering: A top-down approach to {AI} transparency,'' \emph{Computing Research Repository, arXiv Preprint, \textup{arXiv:2310.01405}}, Mar 2025.

\bibitem{Cho_2025}
D.-H. Cho, H.-S. Oh, S.-B. Kim, and S.-W. Lee, ``Emosphere++: Emotion-controllable zero-shot text-to-speech via emotion-adaptive spherical vector,'' \emph{IEEE Transactions on Affective Computing}, vol.~16, no.~3, p. 2365–2380, Jul. 2025.

\bibitem{xie2025emosteerttsfinegrainedtrainingfreeemotioncontrollable}
T.~Xie, S.~Yang, C.~Li, D.~Yu, and L.~Liu, ``Emosteer-tts: Fine-grained and training-free emotion-controllable text-to-speech via activation steering,'' \emph{Computing Research Repository, arXiv Preprint, \textup{arXiv:2508.03543}}, Aug 2025.

\bibitem{qwen3}
A.~Yang, A.~Li, B.~Yang, B.~Zhang, B.~Hui \emph{et~al.}, ``Qwen3 technical report,'' \emph{Computing Research Repository, arXiv Preprint, \textup{arXiv:2505.09388}}, May,2025.

\bibitem{he-etal-2025-sae}
Z.~He, M.~Jin, B.~Shen, A.~Payani, Y.~Zhang, and M.~Du, ``{SAE}-{SSV}: Supervised steering in sparse representation spaces for reliable control of language models,'' in \emph{Proceedings of the 2025 Conference on Empirical Methods in Natural Language Processing}, Suzhou, Jiangsu, China, Nov. 2025, pp. 2207--2236.

\bibitem{zhong2024englishcentricllmslanguagemultilingual}
\BIBentryALTinterwordspacing
C.~Zhong, F.~Cheng, Q.~Liu, J.~Jiang, Z.~Wan, C.~Chu, Y.~Murawaki, and S.~Kurohashi, ``Beyond english-centric llms: What language do multilingual language models think in?'' 2024. [Online]. Available: \url{https://arxiv.org/abs/2408.10811}
\BIBentrySTDinterwordspacing

\bibitem{sun-etal-2025-personality}
S.~Sun, S.~Y. Baek, and J.~H. Kim, ``{Personality Vector}: Modulating personality of large language models by model merging,'' in \emph{Proceedings of the 2025 Conference on Empirical Methods in Natural Language Processing}, Suzhou, Jiangsu, China, Nov. 2025, pp. 24\,656--24\,677.

\bibitem{tang2026facetlevelpersonacontroltraitactivated}
W.~Tang, Z.~Wan, T.~Komamizu, and I.~Ide, ``Facet-level persona control by trait-activated routing with contrastive sae for role-playing llms,'' \emph{Computing Research Repository, arXiv Preprint, \textup{arXiv:2602.19157}}, 2026.

\bibitem{tett2003personality}
R.~P. Tett and D.~D. Burnett, ``A personality trait-based interactionist model of job performance,'' \emph{Journal of Applied Psychology}, vol.~88, no.~3, pp. 500--517, 2003.

\bibitem{zhou2022emotionalvoiceconversiontheory}
K.~Zhou, B.~Sisman, R.~Liu, and H.~Li, ``Emotional voice conversion: Theory, databases and {ESD},'' \emph{Speech Communication}, vol. 137, pp. 1--18, 2022.

\bibitem{6849440}
H.~Cao, D.~G. Cooper, M.~K. Keutmann, R.~C. Gur, A.~Nenkova, and R.~Verma, ``{CREMA-D: Crowd-Sourced Emotional Multimodal Actors Dataset},'' \emph{IEEE Transactions on Affective Computing}, vol.~5, no.~4, pp. 377--390, 2014.

\bibitem{xu-etal-2018-emo2vec}
P.~Xu, A.~Madotto, C.-S. Wu, J.~H. Park, and P.~Fung, ``{E}mo2{V}ec: Learning generalized emotion representation by multi-task training,'' in \emph{Proceedings of the 9th Workshop on Computational Approaches to Subjectivity, Sentiment and Social Media Analysis}, Brussels, Belgium, Oct. 2018, pp. 292--298.

\bibitem{10.1109/TASLP.2022.3164181}
T.~Li, X.~Wang, Q.~Xie, Z.~Wang, and L.~Xie, ``Cross-speaker emotion disentangling and transfer for end-to-end speech synthesis,'' \emph{IEEE/ACM Trans. Audio, Speech and Lang. Proc.}, vol.~30, p. 1448–1460, Apr. 2022.

\bibitem{xu2025qwen25omnitechnicalreport}
J.~Xu, Z.~Guo, J.~He, H.~Hu, T.~He \emph{et~al.}, ``Qwen2.5-omni technical report,'' \emph{Computing Research Repository, arXiv Preprint, \textup{arXiv:2503.20215}}, Mar. 2025.

\bibitem{fang-etal-2025-llama}
Q.~Fang, Y.~Zhou, S.~Guo, S.~Zhang, and Y.~Feng, ``{LL}a{MA}-omni 2: {LLM}-based real-time spoken chatbot with autoregressive streaming speech synthesis,'' in \emph{Proceedings of the 63rd Annual Meeting of the Association for Computational Linguistics}, vol.~1, Vienna, Austria, Jul. 2025, pp. 18\,617--18\,629.

\bibitem{openai2024gpt4o}
J.~Xu, Z.~Guo, J.~He, H.~Hu, T.~He \emph{et~al.}, ``Gpt-4o system card,'' \emph{Computing Research Repository, arXiv Preprint, \textup{arXiv:2410.21276}}, Oct. 2024.

\bibitem{alibabacloud2026}
{Alibaba Cloud}, ``Alibaba cloud documentation,'' \url{https://www.alibabacloud.com/}, 2026, accessed: 2026-02-22.

\bibitem{comanici2025gemini25pushingfrontier}
G.~Comanici, E.~Bieber, M.~Schaekermann, I.~Pasupat, N.~Sachdeva \emph{et~al.}, ``Gemini 2.5: Pushing the frontier with advanced reasoning, multimodality, long context, and next generation agentic capabilities,'' \emph{Computing Research Repository, arXiv Preprint, \textup{arXiv:2507.06261}}, Jul. 2025.

\bibitem{Chen_2022}
S.~Chen, C.~Wang, Z.~Chen, Y.~Wu, S.~Liu \emph{et~al.}, ``Wavlm: Large-scale self-supervised pre-training for full stack speech processing,'' \emph{IEEE Journal of Selected Topics in Signal Processing}, vol.~16, no.~6, p. 1505–1518, Oct. 2022.

\bibitem{ma-etal-2024-emotion2vec}
\BIBentryALTinterwordspacing
Z.~Ma, Z.~Zheng, J.~Ye, J.~Li, Z.~Gao \emph{et~al.}, ``emotion2vec: Self-supervised pre-training for speech emotion representation,'' in \emph{Findings of the Association for Computational Linguistics: ACL 2024}, L.-W. Ku, A.~Martins, and V.~Srikumar, Eds.\hskip 1em plus 0.5em minus 0.4em\relax Bangkok, Thailand: Association for Computational Linguistics, Aug. 2024, pp. 15\,747--15\,760. [Online]. Available: \url{https://aclanthology.org/2024.findings-acl.931/}
\BIBentrySTDinterwordspacing

\bibitem{du2025cosyvoice3inthewildspeech}
\BIBentryALTinterwordspacing
Z.~Du, C.~Gao, Y.~Wang, F.~Yu, T.~Zhao \emph{et~al.}, ``Cosyvoice 3: Towards in-the-wild speech generation via scaling-up and post-training,'' 2025. [Online]. Available: \url{https://arxiv.org/abs/2505.17589}
\BIBentrySTDinterwordspacing

\end{thebibliography}

\end{document}